\documentclass[12pt,preprint]{aastex}
\usepackage{epsfig}

\newcommand{\bk}{ \mbox{\boldmath$k$} }
\newcommand{\bx}{{\mbox{\boldmath$x$}}}
\newcommand{\br}{{\mbox{\boldmath$r$}}}
\newcommand{\rhat}{ \mbox{\boldmath$\hat{r}$} }
\newcommand{\thetahat}{ \mbox{\boldmath$\hat{\theta}$} }
\newcommand{\zhat}{ \mbox{\boldmath$\hat{z}$} }
\newcommand{\bbv}{ \mbox{\boldmath$v$} }
\newcommand{\bB}{ \mbox{\boldmath$B$} }
\newcommand{\bnabla}{ \mbox{\boldmath$\nabla$} }
\newcommand{\curl}{ \mbox{\boldmath$\times$} }

\begin{document}
\shorttitle{Scattering of acoustic waves by a magnetic cylinder}
\shortauthors{Gizon, Hanasoge, \& Birch}
\title{Scattering of acoustic waves by a magnetic cylinder: \\
 Accuracy of the Born approximation}
\author{L. Gizon}
\affil{Max-Planck-Institut f\"{u}r Sonnensystemforschung, 37191 Katlenburg-Lindau, Germany}
\author{S.~M. Hanasoge}
\affil{W.~W. Hansen Experimental Physics Laboratory, Stanford University, Stanford, CA 94305}
\author{A.~C. Birch}
\affil{Colorado Research Associates Division, NorthWest Research Associates Inc., 3380 Mitchell Lane, Boulder, CO 80301}

\begin{abstract}
With the aim of studying magnetic effects in time-distance helioseismology, we use the first-order Born approximation to compute the scattering of acoustic plane waves by a magnetic cylinder embedded in a uniform medium. We show, by comparison with the exact solution, that the travel-time shifts computed in the Born approximation  are  everywhere valid to first order in the ratio of the magnetic to the gas pressures.  We also show that, for arbitrary magnetic field strength, the Born approximation is not valid in the limit where the radius of the magnetic cylinder tends to zero. 
\end{abstract}


\section{Introduction}
Time-distance helioseismology \citep{Duvall1993} has been used to measure wave travel times in and around magnetic active regions and sunspots to estimate subsurface flows and wave-speed perturbations \citep[e.g.][]{Duvall1996, Kosovichev2000}.  A challenging problem is to estimate the subsurface magnetic field from travel times. In order to do so, one must understand the dependence of the travel times on the magnetic field. 

As discussed by e.g.\ \citet{Cally2005} the interaction of acoustic waves with sunspot magnetic fields is strong in the near surface layers.  As a result, the effect of the magnetic field on the travel times is not expected to be small near the surface. Deeper inside the Sun, however, the ratio of the magnetic pressure to the gas pressure becomes small, and it is tempting to treat the effects of the magnetic field on the waves using perturbation theory. The hope is to eventually develop a linear inversion to estimate the subsurface magnetic field from travel times measured between surface locations that are free of magnetic field. Of particular interest is the search for a magnetic field at the bottom of the convection zone. Such a linear inversion scheme has been proposed by \citet{Kosovichev1997} for time-distance helioseismology using the ray approximation, but it needs to be extended to finite wavelengths.

As a first step, in this paper, we consider the scattering of small amplitude acoustic plane waves by a magnetic cylinder embedded in a uniform medium. This simple problem has a known exact solution for arbitrary magnetic field strengths \citep{Wilson1980}. The first-order Born and Rytov approximations have proved useful in the context of time-distance helioseismology to model the effects of small local perturbations in sound speed and flows \citep[e.g.][]{Birch2001,Jensen2003, Birch2004}. Here we use the Born approximation to compute the scattering of a wave by a weak magnetic field. The validity of the Born approximation is not a priori obvious in this case, since the magnetic field allows additional wave modes. Because we have an exact solution, however, we can study the validity of the linearization of travel times on the square of the magnetic field. We note that the problem of the scattering of waves by a non-magnetic cylinder with a sound speed that differs from the surrounding medium was investigated by \citet{Fan1995}.

The outline of the paper is as follows.  In Section~\ref{sec.the_problem} we specify the problem and write the equations of motion for small amplitude waves. In Section~\ref{sec.exact_solution} we review the exact solution to the scattering problem.  In Section~\ref{sec.born} we apply the first Born approximation to obtain the complex scattering amplitudes.  In Section~\ref{sec.compare} we show that the Born approximation is an asymptote of the exact solution in the limit of infinitesimal magnetic field strength. In Section~\ref{sec.travel_times} we compare travel times computed exactly, in the Born approximation, and in the ray approximation. In Section~\ref{sec.discussion} we provide a brief summary of our results and also discuss the limit when the magnetic tube radius tends to zero.

\section{The Problem}
\label{sec.the_problem}
\subsection{Governing equations}
We start with the ideal equations of magnetohydrodynamics. The equations of continuity, momentum, magnetic induction, and Gauss' law for the magnetic field are:
\begin{eqnarray}
D_t  \rho + \rho \bnabla \cdot \bbv &=& 0  , \label{eq.first} \\
  \rho D_t \bbv + \bnabla p - \frac{1}{4\pi} (\bnabla \curl \bB) \curl \bB  &=&0 \label{mom_init}  , \\ 
 \partial_t \bB  - \bnabla \curl ( \bbv \curl \bB)  &=& 0  , \\
\bnabla\cdot\bB &=&0  ,  
\end{eqnarray}
where  $D_t = \partial_t + \bbv\cdot\bnabla$ is the material derivative, $\rho$ the density, $\bbv$ the velocity, $p$ the pressure, and $\bB$ the magnetic field.  For the sake of simplicity, we use the simple energy equation
\begin{equation}
\rho C_v   D_t T   + p \bnabla \cdot \bbv  = 0 ,
\label{energy1}
\end{equation}
where $T$ is the temperature, $C_v=R/(\gamma-1)$ the uniform specific heat at constant volume, $R$ the gas constant, and $\gamma$ the ratio of specific heats. This equation neglects all forms of heat losses. In addition we use the ideal gas equation of state
\begin{equation}
p = \rho R T  . \label{eq.last}
\end{equation}

\subsection{Steady background state}
We consider a magnetic cylinder with radius $R$ and uniform magnetic field strength $B_t$ embedded in an infinite, otherwise uniform, gravity free medium with constant density $\rho_0$, gas pressure $p_0$, and temperature $T_0$. We use a cylindrical coordinate system $(r,\theta,z)$ where $r$ is the radial coordinate, $\theta$ is the azimuthal angle, and $z$ is the vertical coordinate in the direction of the cylinder axis. We denote the corresponding unit vectors by $\rhat$, $\thetahat$, and $\zhat$. All steady physical quantities are denoted with an overbar. In particular, we have
\begin{eqnarray}
\overline{\bB} &=& B_t \Theta(R-r) \zhat  , \\
\overline{\rho} &=& \rho_t \Theta(R-r)  + \rho_0 \Theta(r-R)  , \\
\overline{p} &=& p_t \Theta(R-r)  + p_0 \Theta(r-R)  ,
\end{eqnarray}
where the Heaviside step function is defined by $\Theta(r) = 0$ if $r < 0$ and $\Theta(r) = 1$ if $r > 0$. The density and pressure inside the tube are $\rho_t$ and $p_t$ respectively. We assume that there is no mean flow in this problem, i.e.\ $\overline{\bbv}= 0$.

We choose to study the case where the background temperature is the same inside and outside the magnetized region, i.e. $\overline{T} = T_t = T_0$. As a result,  the sound speed, $c = ( \gamma R \overline{T} )^{1/2}$, is constant everywhere. This choice is motivated by our desire to restrict ourselves, as much as possible, to the study of the effect of the Lorentz force on waves, rather than the effect of a sound speed variation. Pressure balance across the magnetic tube boundary implies
\begin{equation}
p_t + {B_t^2}/{8\pi}  = p_0  , \label{eq.pressure_bal}
\end{equation} 
where $p_t$ is the background gas pressure inside the tube. The density inside the tube is given by $\rho_t = \rho_0 p_t/p_0$, as the temperature is the same inside and outside the tube. 

\subsection{Linear waves}
We want to study the propagation of linear waves on the steady background state defined above. Toward this end, we expand each physical quantity that appears in equations~(\ref{eq.first})-(\ref{eq.last}) into a time-varying component,  denoted by a prime, and the steady component, denoted with an overbar. For example, we write $p=\overline{p}+p'$. After subtraction of the steady state, we obtain:
\begin{eqnarray}
\partial_t\rho' &=& - \bnabla\cdot(\overline{\rho} \bbv') \label{cont} , \\
\overline\rho\partial_t\bbv' + \bnabla p' &=&   \frac{1}{4\pi} \left[(\bnabla\curl\bB')\curl\overline{\bB} + (\bnabla\curl\overline{\bB})\curl\bB'\right] ,
 \label{mom}  \\
\partial_t\bB' &=&   \bnabla\curl\left(\bbv'\curl\overline{\bB}\right) , \\
\bnabla\cdot\bB' &=& 0 .    
\end{eqnarray}
The linearized energy equation, in combination with equation~(\ref{cont}) and the linearized equation of state,  may be simplified to 
\begin{equation}
\partial_t p' - c^2\partial_t\rho' = \frac{\gamma-1}{\gamma} c^2 \bbv' \cdot\bnabla\overline{\rho}  , \label{temp}
\end{equation}
which describes adiabatic wave motion.

As we study linear waves on a steady background, we can consider one temporal Fourier mode at a time. The magnetic field $\overline{\bB}$ and all other background quantities do not depend on $z$. Thus, a wave with a $z$ dependence of the form $e^{i k_z z}$ will have the same $z$ dependence after interacting with the magnetic cylinder.  As a result, we study solutions where the pressure fluctuations are of the form
\begin{equation}
p'(\br,z,t) = \tilde{p}(\br) \exp(i k_z z - i \omega t)  ,  \label{eq.expand}
\end{equation}
where $\br=(r,\theta)$ is a position vector perpendicular to the tube axis. All the other wave variables, $\rho'$, $\bbv'$, and $\bB'$ are written in the same form as equation~(\ref{eq.expand}). Quantities with a tilde only depend on $\br$.

\section{Exact Solution}
\label{sec.exact_solution}
For the sake of completeness, we briefly review an exact solution obtained by \citet{Wilson1980} to equations~(\ref{cont})-(\ref{temp}). We consider a plane wave incident on the magnetic tube, with pressure fluctuations of the form 
\begin{equation}
\tilde{p}_{\rm inc}(\br) = P \exp(i\bk\cdot\br)   ,
\label{eq.definc}
\end{equation}
 where $P$ is an amplitude and $\bk$ is the component of the wave vector perpendicular to the tube axis.
In order for the incident wave to be a solution to the non-magnetic problem, the horizontal wavenumber $k=\|\bk\|$ must satisfy
\begin{equation}
k(\omega) = \sqrt{\omega^2/c^2-k_z^2} \, .
\end{equation}
In the rest of this paper, unless otherwise stated, we will use $k$ to denote $k(\omega)$.
In cylindrical coordinates, this plane wave can be expanded as a sum over azimuthal components (index $m$) according to \citep[e.g.][]{Bogdan1989}:
\begin{equation}
\tilde{p}_{\rm inc}(\br) =  P \sum_{m = -\infty}^{\infty} {i}^m J_m(kr) e^{i m\phi }  ,
\label{incident}
\end{equation}
where $J_m$ denotes the Bessel function of order $m$ and $\phi$ is the angle between $\bk$ and $\br$. 

The total wave pressure, hydrodynamic plus magnetic, and the radial velocity must be continuous across the tube boundary.
Applying these boundary conditions, \citet{Wilson1980} showed that the total pressure wave field is
\begin{equation}
\label{eq.exact_solution}
\tilde{p}(\br) = \left\{ \begin{array}{ll}
P \sum_m  i^m B_m J_m(k_tr) e^{i m\phi}  & r<R  \\
\tilde{p}_{\rm inc} + P \sum_m i^m A_{m} H_m(kr) e^{i m\phi} & r>R  
\end{array} \right.
\end{equation}
where $H_m=H_m^{(1)}$ is the Hankel function of the first kind of order $m$. The quantity $k_t$ is the horizontal wavenumber inside the tube, given by
\begin{equation}
k_t = k\left[\frac{(\omega^2 - k_z^2 {a}^2)}{(1 + {a}^2/c^2)(\omega^2 - s^2 k_z^2)}\right]^{1/2}
\label{modified}  ,
\end{equation}
where ${a} = B_t / \sqrt{4\pi\rho_t}$
is the Alfv\`{e}n wave speed and $s = {a} c\left({a}^2 + c^2\right)^{-1/2}$ is the tube velocity. The coefficients $A_m$ and $B_m$ are given in Appendix~\ref{app.exact}. This exact solution is valid for arbitrarily large values of $B_t$ and $R$. We note that the gas pressure fluctuations are discontinuous at the tube boundary.

\begin{figure}
\plotone{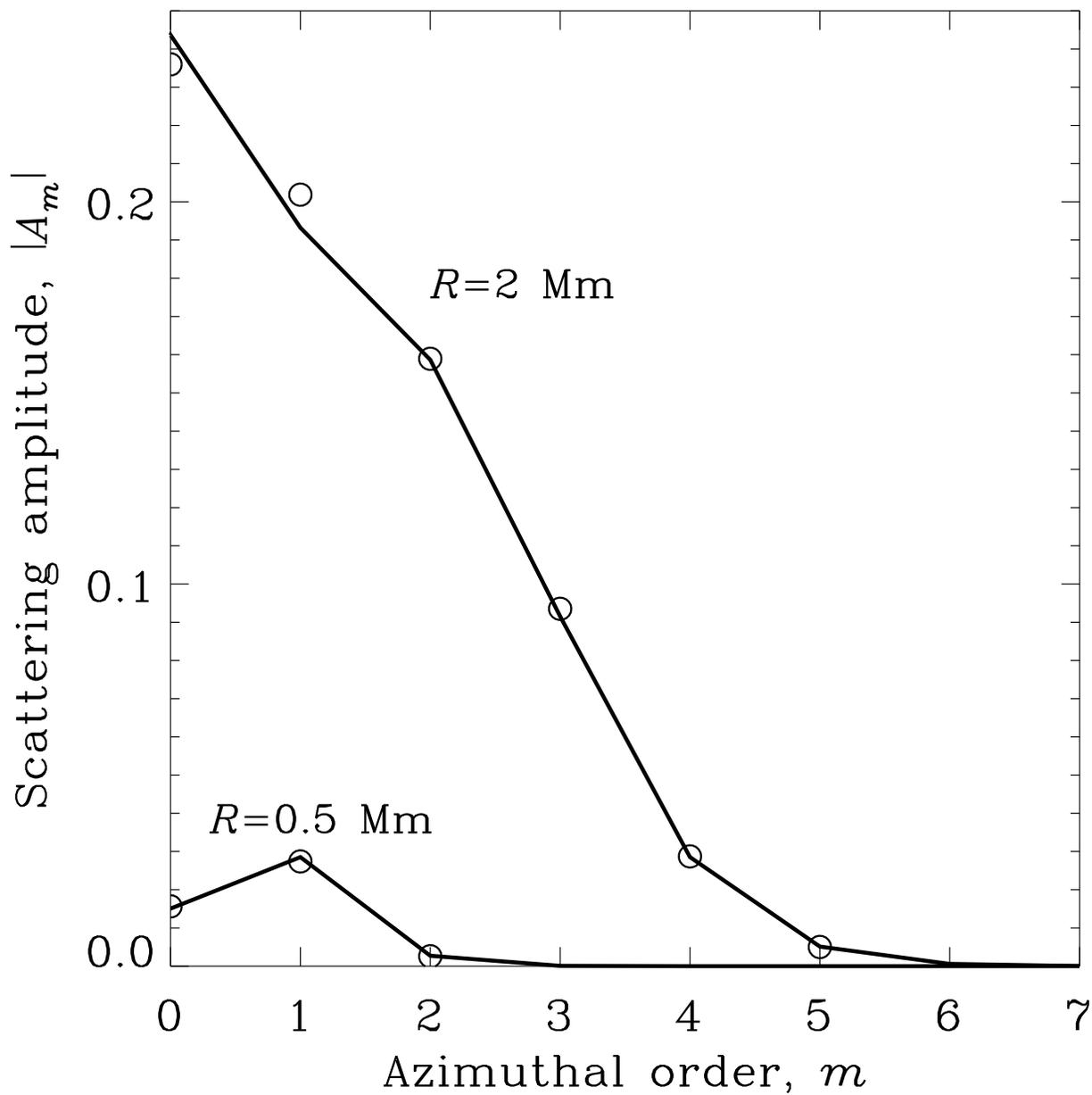}
\caption{Coefficients $|A_m|$ as a function of $m$ for $R=2$~Mm and $R=0.5$~Mm (thick lines). Only the $m>0$ values are shown since $|A_{-m}| = |A_m|$. The magnetic field is $B=1$~kG, $\omega/2\pi=3$~mHz, and $k_z=0$.  The coefficients $A_m$ are negligible for $m>5$ in the case $R=2$~Mm and for $m>2$ in the case $R=0.5$~Mm.  The open circles are the Born approximation to these coefficients ($|A^{\rm  Born}_m|$, see section~\ref{sec.born}). }
\label{fig.exact_and_born}
\end{figure}

Not all azimuthal components $m$ contribute at the same level. Figure~\ref{fig.exact_and_born} shows the amplitudes of the $A_m$ as a function of $m$ for two different values of the tube radius, $R$.  In this particular example,  $\rho_0=5\times10^{-7}$~cgs,  $c=11$~km/s, and $B=1$~kG are fixed. For $R=0.5$~Mm, which is less than the wavelength ($kR=0.86$),  only the $m=0,\pm1$ azimuthal components contribute.  For tubes with larger radii the higher order components have larger amplitudes, as can be seen in the case $R=2$~Mm ($kR = 3.43$).

\section{First Born approximation}
\label{sec.born}
Magnetic effects cause perturbations to both the steady background state and the wavefield. In this section we use the Born approximation  to derive an approximate solution to equations~(\ref{cont})-(\ref{temp}) based on the assumption that magnetic effects are small \citep[see e.g.][]{Rosenthal1995}.  The Lorentz force is quadratic in the magnetic field.  As a result we introduce a small parameter that is second order in the magnetic field.   We choose to expand all quantities   in powers of the small dimensionless parameter
\begin{equation}
\epsilon =  \frac{B_t^2}{4\pi\rho_0 c^2}  \, .
\end{equation}
In this expansion framework  the magnetic field appears at order  $\epsilon^{1/2}$.  In particular, we write the steady background magnetic field as
\begin{equation}
\overline{\bB} = \epsilon^{1/2} \bB_1  .
\end{equation}

The magnetic field causes a shift, $\epsilon\rho_1$, in the steady component of the density inside the tube relative to the steady component of the density outside the tube, $\rho_0$:
\begin{equation}
\overline{\rho} = \rho_0 + \epsilon \rho_1   .
\end{equation}
Likewise, we write the steady component of pressure as
\begin{equation}
\overline{p} = p_0 + \epsilon p_1  .
\end{equation}
These changes are related to the magnetic field through equation~(\ref{eq.pressure_bal}) and the equation of state:
\begin{eqnarray}
p_1  &=& - \frac{\rho_0 c^2}{2} \Theta(R-r)    , \\
\rho_1 &=& - \frac{\gamma\rho_0}{2} \Theta(R-r)   .
\end{eqnarray}

We expand each of the wave-field variables into an incident component (subscript ``inc'') and a scattered component (subscript ``sc''):
\begin{eqnarray}
p' &=& p'_{\rm inc} + \epsilon p'_{\rm sc}  , \label{eq.totalp} \\
\rho' &=& \rho'_{\rm inc} + \epsilon \rho'_{\rm sc}  , \\
\bbv' &=&  \bbv'_{\rm inc} + \epsilon \bbv'_{\rm sc}    , \\
\bB' &=&  \epsilon^{1/2} \bB'_{\rm sc}    . 
\end{eqnarray}
By inserting the above expansions into equations~(\ref{cont})-(\ref{temp}) and retaining the terms of order $\epsilon$ we obtain
\begin{eqnarray}
-i\omega\rho'_{\rm sc} + \rho_0 \bnabla \cdot \bbv'_{\rm sc} &=& - \bnabla \cdot (\rho_1\bbv'_{\rm inc}) , \\
-i\omega \rho_0\bbv'_{\rm sc} + \bnabla p'_{\rm sc} &=&  i\omega\rho_1\bbv'_{\rm inc} + \frac{1}{4\pi} (\bnabla \curl \bB_1) \curl {\bB}'_{\rm sc}  \nonumber  \\
&& + \frac{1}{4\pi} (\bnabla \curl {\bB}'_{\rm sc}) \curl \bB_1  , \\
-i\omega \left({p}'_{\rm sc} - c^2 {\rho}'_{\rm sc}\right) &=& \frac{(\gamma - 1)c^2}{\gamma}  {\bbv}'_{\rm inc} \cdot \bnabla \rho_1 , \\
-i\omega{\bB}'_{\rm sc} &=& \bnabla \curl ({\bbv}'_{\rm inc} \curl \bB_1) .
\end{eqnarray}
The terms on the right-hand side of the above equations act as sources for the scattered waves: this is the Born approximation.
Writing all wave variables in the form of equation~(\ref{eq.expand}) and
using the fact that the magnetic field is solenoidal, the above equations reduce to a forced Helmholtz equation for the $(k_z, \omega)$ Fourier component  of the scattered pressure field, $\tilde{p}_{\rm sc}$:
\begin{equation}
\left( \Delta_\br + k^2 \right) \tilde{p}_{\rm sc}(\br)  =  \tilde{S}(\br)   ,
\label{diffeqn}
\end{equation}
where $\Delta_\br$ is the two-dimensional Laplacian with respect to $\br$ and the source function $\tilde{S}(\br)$ is given by
\begin{eqnarray}
\tilde{S}(\br) &=& \frac{\gamma-1}{2} \delta(r-R) \partial_r \tilde{p}_{\rm inc} (\br)  \nonumber \\
&& - \frac{c^2 k^2}{\omega^2}  (\Delta_\br - k_z^2)[ \Theta(r-R) \tilde{p}_{\rm inc} (\br)]  \nonumber \\
 && -  \frac{c^2}{2\omega^2} (\Delta_\br - 3 k_z^2) \left[  \delta(r-R) \partial_r \tilde{p}_{\rm inc} (\br) \right] .
\label{source} 
\end{eqnarray}
The first term in $\tilde{S}$ is due to the density jump at the tube boundary and the other two terms are due to the direct effect of the Lorentz force on the wave.  For the incoming wave, $\tilde{p}_{\rm inc}$, we take the same plane wave as in the exact solution (Eq.~[\ref{eq.definc}]).

\begin{figure*}
\plottwo{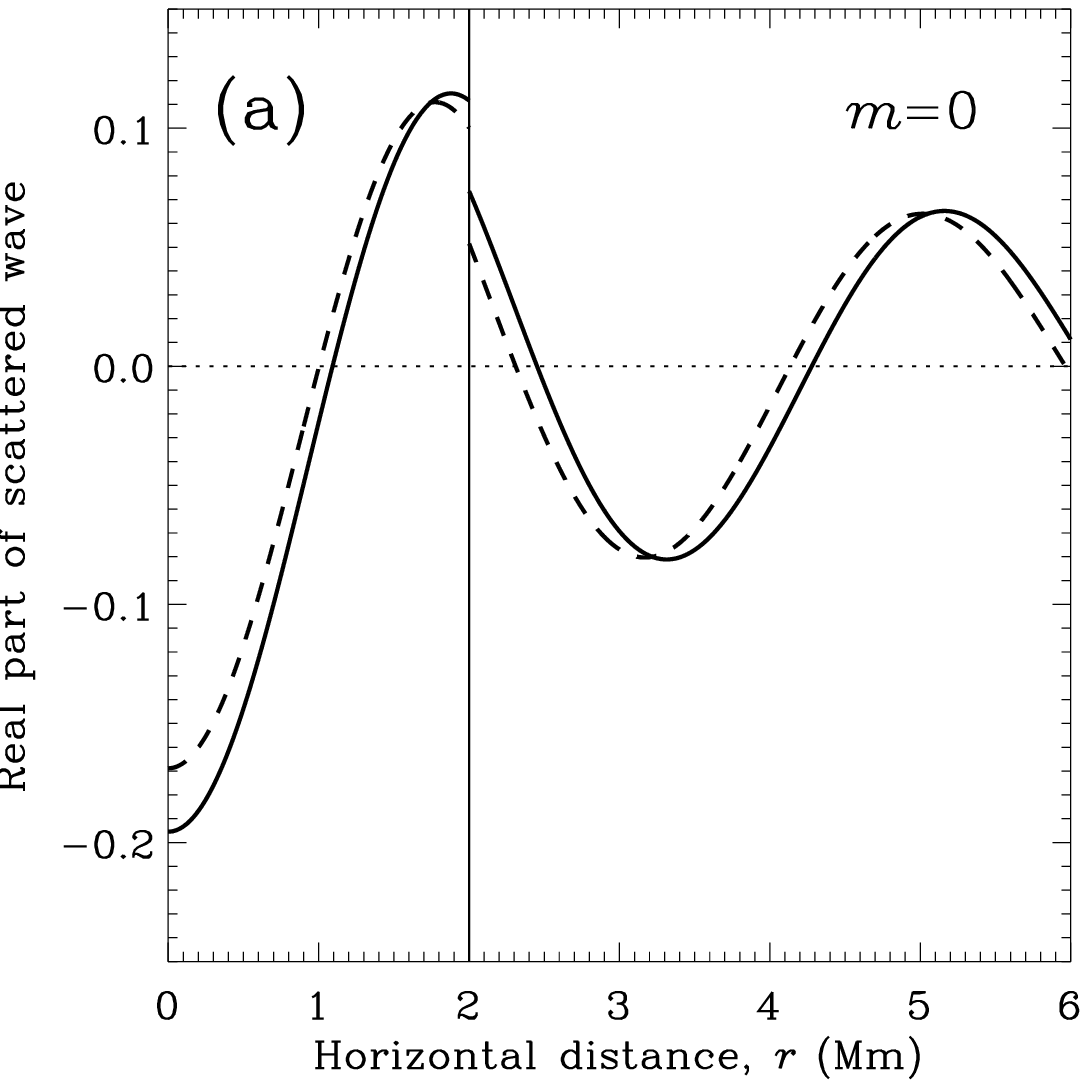}{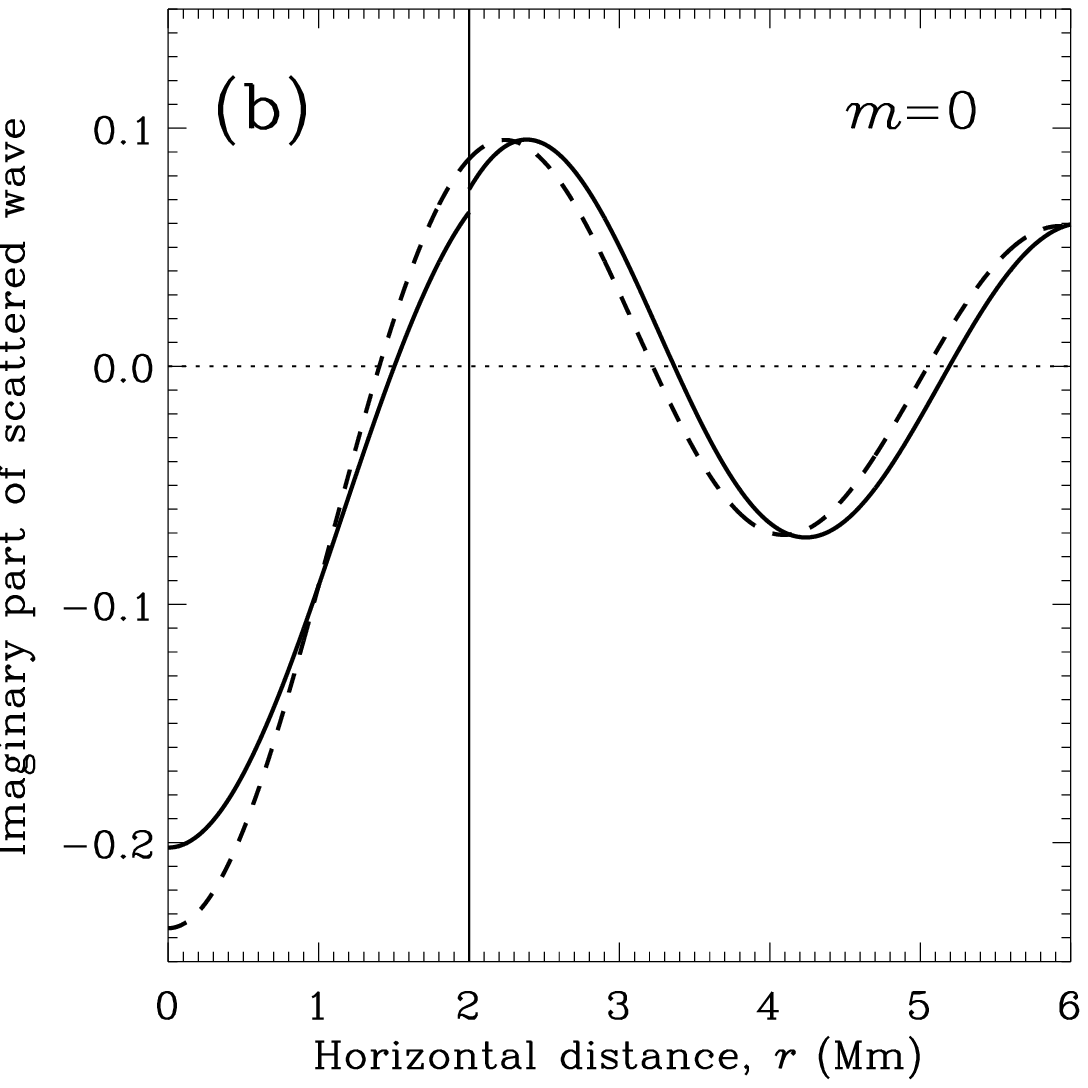}
\caption{ a) Real and b) imaginary parts of the scattered pressure field in the Born approximation (dashed line) and the exact solution (solid line).  In this case the incoming wave is of the form $\tilde{p}_{\rm inc}=J_0(kr)$ ($m=0$) and we used $B=1$~kG, $R=2$~Mm, $k_z=0$, and $\omega/2\pi=3$~mHz.}
\label{fig.real_imag}
\end{figure*}

The solution to the inhomogeneous Helmholtz equation~(\ref{diffeqn}) is
\begin{equation}
\tilde{p}_{\rm sc}(\br) = \int\!\!\!\int G(\br|\br') \tilde{S}(\br') \,  d\br' 
\label{master}
\end{equation}
where  $G(\br|\br')$ is the Green's function defined by
\begin{equation}
\left(\Delta_{\br} + k^2\right) G(\br|\br') = \delta(\br-\br')  
\label{green}
\end{equation}
and explicitly given by \citep[e.g.][]{Morse1986} 
\begin{equation}
G(\br|\br') 
= - \frac{i}{4} \sum_{m=-{\infty}}^{\infty}  H_m (k r_>) J_m(kr_<)e^{im(\theta-\theta')}  ,
\label{greenfunc}
\end{equation}
where $\br=(r,\theta)$, $\br'=(r',\theta')$, and $r_> = {\rm max}(r, r')$ and $r_< = {\rm min}(r, r')$. 

We insert this expression for $G$ into equation~(\ref{master}) and use integration by parts as appropriate. The solution can be written in terms of integrals over bilinear combinations of Bessel and Hankel functions. These integrals can be evaluated using equations~(\ref{eq.appendix1}) and~(\ref{eq.appendix2}). Upon simplification, we find that the pressure of the scattered wave is
\begin{eqnarray}
\epsilon\tilde{p}_{\rm sc} (\br) &=& P \sum_{m=-\infty}^{\infty} i^m e^{i m\phi}
\nonumber \\
&& \times
\left\{
\begin{array}{ll}
   C_m J_m(kr) -\epsilon\frac{kr}{2}  J'_m(kr) 
&  \; r < R   \\
  A^{\rm Born}_m H_m(kr) 
& \; r > R  , 
\end{array}
\right. 
\end{eqnarray}
where the coefficients $A^{\rm Born}_m$ and $C_m$ are given in Appendix~\ref{app.Born}. 

Figure~\ref{fig.real_imag} shows the real and imaginary parts of the Born approximation for the scattered pressure field, $\epsilon\tilde{p}_{\rm sc}$, for the case $\tilde{p}_{\rm inc}=J_0(kr)$, i.e. when $m=0$ and $P=1$. In this example, $\epsilon=0.13$. On the same figure, we also show the exact calculation of the scattered pressure field obtained by subtracting $\tilde{p}_{\rm inc}$ from the complete exact pressure field, $\tilde{p}$, computed as in Section~\ref{sec.exact_solution}.  The amplitude of the scattered wave is of order $\epsilon$ of the amplitude of the incoming wave, as expected.  The Born approximation is everywhere accurate to about 10\%.  

For a direct comparison with the exact solution (Eq.~[\ref{eq.exact_solution}]), we remind the reader that the total pressure wave field in the Born approximation is given by $\tilde{p} = \tilde{p}_{\rm inc} + \epsilon \tilde{p}_{\rm sc}$, according to equation~(\ref{eq.totalp}).

\section{Born tends to  the exact solution as $\epsilon\rightarrow 0$}
\label{sec.compare}
\begin{figure}
\plotone{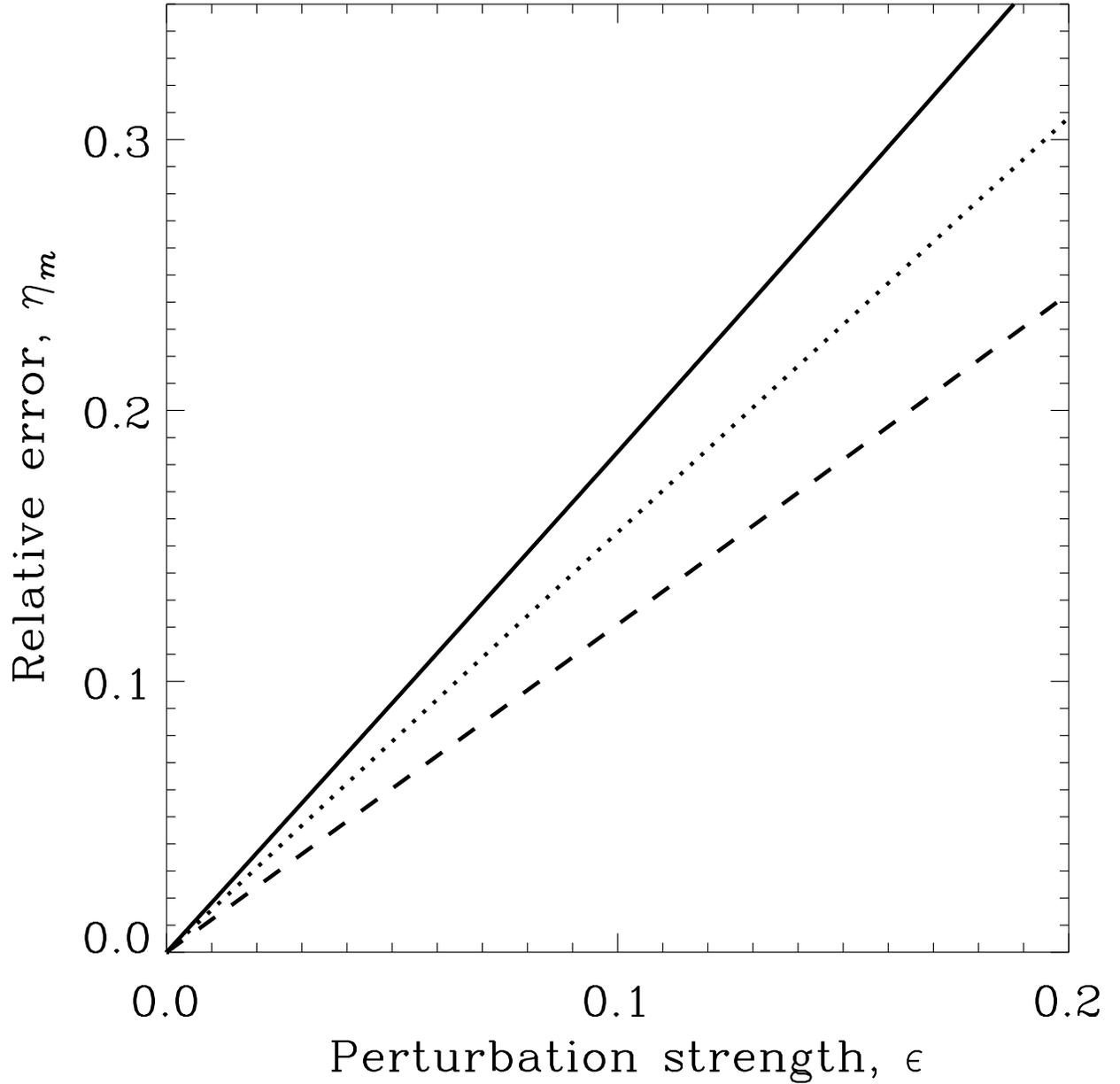}
\caption{The fractional error $\eta_m= | A^{\rm Born}_m - A_m| / |A_m|$ as a function of $\epsilon$ for $m=0$ (solid line), $m=1$ (dotted line), and $m=2$ (dashed line) for the case $R=2$~Mm, $\omega/2\pi=3$~mHz, and $k_z=0$.  }
\label{fig.vary_epsilon}
\end{figure}

In this section we show that to first order in $\epsilon$, the exact solution (Sec.~\ref{sec.exact_solution}) and the Born solution (Sec.~\ref{sec.born})  are identical. We expand the exact solution (Eqs.[\ref{eq.exact_solution}], [\ref{exact_out}], and [\ref{exact_in}]) in a Taylor series up to first order in $\epsilon$. To do this, we use
\begin{equation}
\frac{a^2}{c^2} = \frac{\epsilon}{1-\gamma\epsilon/2}  
\end{equation}
which, together with equation~(\ref{modified}), gives the first-order perturbation to the wavenumber inside the tube,
\begin{equation}
k_t(\epsilon)  =  k (1- \epsilon/2 ) + {\rm O}(\epsilon^2)   .
\end{equation}
Let us first consider the exact scattering coefficient $A_m$ outside the tube ($r>R$) given by equation~(\ref{exact_out}). Denoting the numerator and denominator of $A_m$ by $N$ and $D$ respectively and performing a Taylor expansion, we obtain 
\begin{eqnarray}
N &=& \frac{1}{2} \epsilon \left[(\gamma + 2\frac{c^2k_z^2}{\omega^2}) J'_m(kR) J_m(kR) + kR J^2_m(kR)
\right. \nonumber \\
&& \left. 
 - kR J_{m-1}(kR)J_{m+1}(kR) \right] + {\rm O}(\epsilon^2) \label{second}
\end{eqnarray}
and
\begin{eqnarray}
D &=& J_m(kR)H'_m(kR) - J'_m(kR)H_m(kR) + {\rm O}(\epsilon) 
\nonumber \\
&=& \frac{2i}{\pi kR} + {\rm O}(\epsilon) .
\label{third}
\end{eqnarray}
Hence, the exact and Born coefficients outside the tube match to first order in $\epsilon$: 
\begin{equation}
A_m = A^{\rm Born}_m + {\rm O}(\epsilon^2)   .
\label{eq.limit_out}
\end{equation}
Similarly, it can be demonstrated that the coefficient $B_m$ that gives the exact total wave field inside the tube ($r<R$) is
\begin{equation}
B_m - 1 = C_m  + {\rm O}(\epsilon^2)   .
\label{eq.limit_in}
\end{equation}
The minus one on the left-hand side comes from the fact that the $B_m$ coefficient relates to the full wavefield, whereas $C_m$ is for the scattered wavefield only. Together, equations~(\ref{eq.limit_out}) and~(\ref{eq.limit_in}) imply that the Born approximation is identical to the exact solution outside and inside the magnetic tube, to first order in $\epsilon$. Figure~\ref{fig.vary_epsilon} shows the  fractional error $\eta_m = | A^{\rm Born}_m - A_m| / |A_m|$ as a function of $\epsilon$ for $m=0$, $m=1$, and $m=2$.  We see that the fractional error of the Born approximation tends to zero as  $\epsilon$ tends to zero.

\section{Travel Times}
\label{sec.travel_times}
In this section, we study the interaction of solar-like wave packets with the magnetic cylinder. The aim is to compare seismic travel-time shifts computed in the Born approximation and from the exact solution. For the sake of simplicity, we fix $k_z=0$. Using Cartesian coordinates $\br=(x,y)$ for the horizontal plane, we choose an incoming Gaussian wavepacket propagating in the $+\bx$ horizontal direction:
\begin{equation}
p'_{\rm inc}(\br,t) =  \int_0^\infty e^{-(\omega-\omega_*)^2/(2\sigma^2)}  \cos[ k(\omega) x  - \omega t ]  \,  d\omega  .
\end{equation}
where $\omega_*/2\pi = 3$~mHz is the dominant frequency of solar oscillations and $\sigma/2\pi = 1$~mHz is the dispersion. Since we chose $k_z=0$, we have $k(\omega) = \omega/c$. The wavepacket is centered on the magnetic tube at time $t=0$. The scattered wave packet can be calculated, exactly or in the Born approximation, from the previous sections.

\begin{figure*}
\plotone{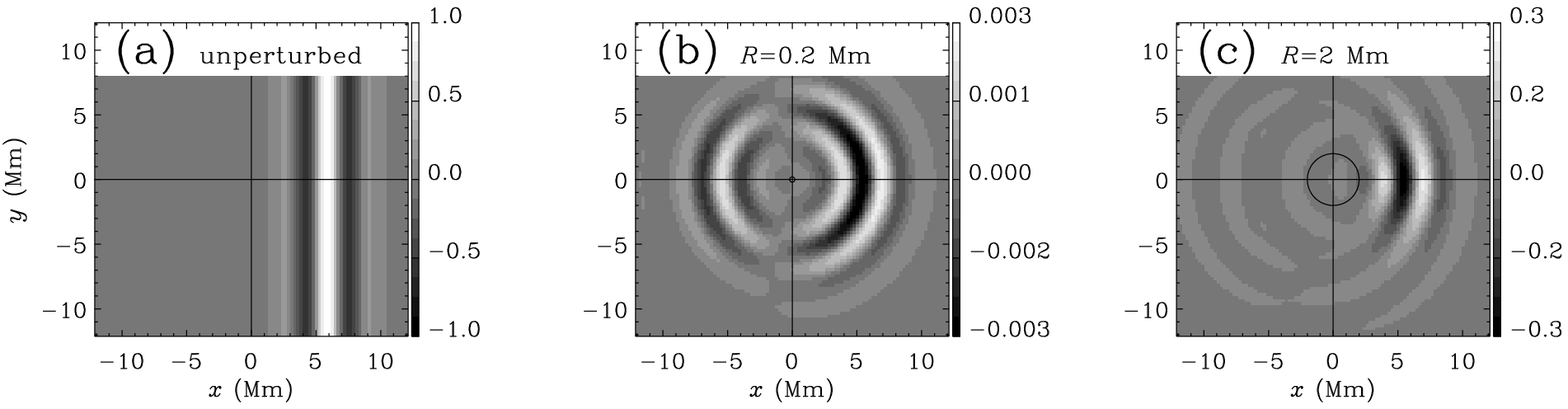}
\caption{Plots of the pressure field of the unperturbed (a) and scattered (b - c) wavepackets  at time $t=8.9$~min after the unperturbed wavepacket has crossed the center of the magnetic cylinder. The wavepacket parameters are described in the text. The wavevector is normal to the axis of the magnetic cylinder and in the $+\hat{\bf x}$ direction. The strength of the magnetic perturbation is $\epsilon=0.13$. In panel (b) the tube radius is $R=0.2$~Mm and in panel (c) is the tube radius is $R=2$~Mm. The circles outline the cross-section of the tube.  Notice that the gray scales are different in each panel. The backscattered wave is more prominent for the tube with the smaller radius (panel b).}
\label{fig.frames}
\end{figure*}

The three panels in Figure~\ref{fig.frames} show snapshots of the incoming and scattered pressure fields at time $t=8.9$~min. The parameters of the steady background at infinity are $\rho_0=5\times10^{-7}$~cgs and $c=11$~km/s, which are roughly the conditions at a depth of $250$~km below the solar photosphere.  The incident wavepacket is shown in Figure~\ref{fig.frames}a. Figure~\ref{fig.frames}b shows the scattered wave which results from a $1$-kG magnetic flux tube with radius $R=0.2$~Mm.  The smaller tube produces relatively more back scattering than the large tube in comparison with the forward scattering. The amplitude of the scattered wave is roughly three orders of magnitude smaller than the incoming wave.  Figure~\ref{fig.frames}c shows the scattered wave for a larger tube radius of $2$~Mm; the scattering is dominantly in the forward direction and has an amplitude only an order of magnitude smaller than the incident wave.

We now define the travel-time shifts that are caused by the magnetic cylinder. By definition, the travel-time shift at location $\br$  is the time $\delta t(\br)$ which minimizes the function
\begin{equation}
X(t) = \int dt' \left[  p'(\br, t') -  p'_{\rm inc}(\br, t'-t) \right]^2   ,
\label{eq.def_tt}
\end{equation}
where $p'$ is the full wavefield that includes both the incident wavepacket and the scattered wave packet caused by the magnetic field. The travel-time shifts can be computed in this way for either the exact solution or the Born-approximation. 

In addition, it is also interesting to compare with the ray approximation as given by equation~(14) from \citet{Kosovichev1997}. In our case, where  $\bk\cdot\overline{\bB}=0$ and the magnetic field strength is constant inside the tube, the ray approximation becomes $\delta t(\br) = -L(\br) a^2/ (2 c^3) $ where $L(\br)$ is the path length through the tube  along the ray which goes from coordinates $(-\infty,y)$ to $\br=(x,y)$.

\begin{figure*}
\plottwo{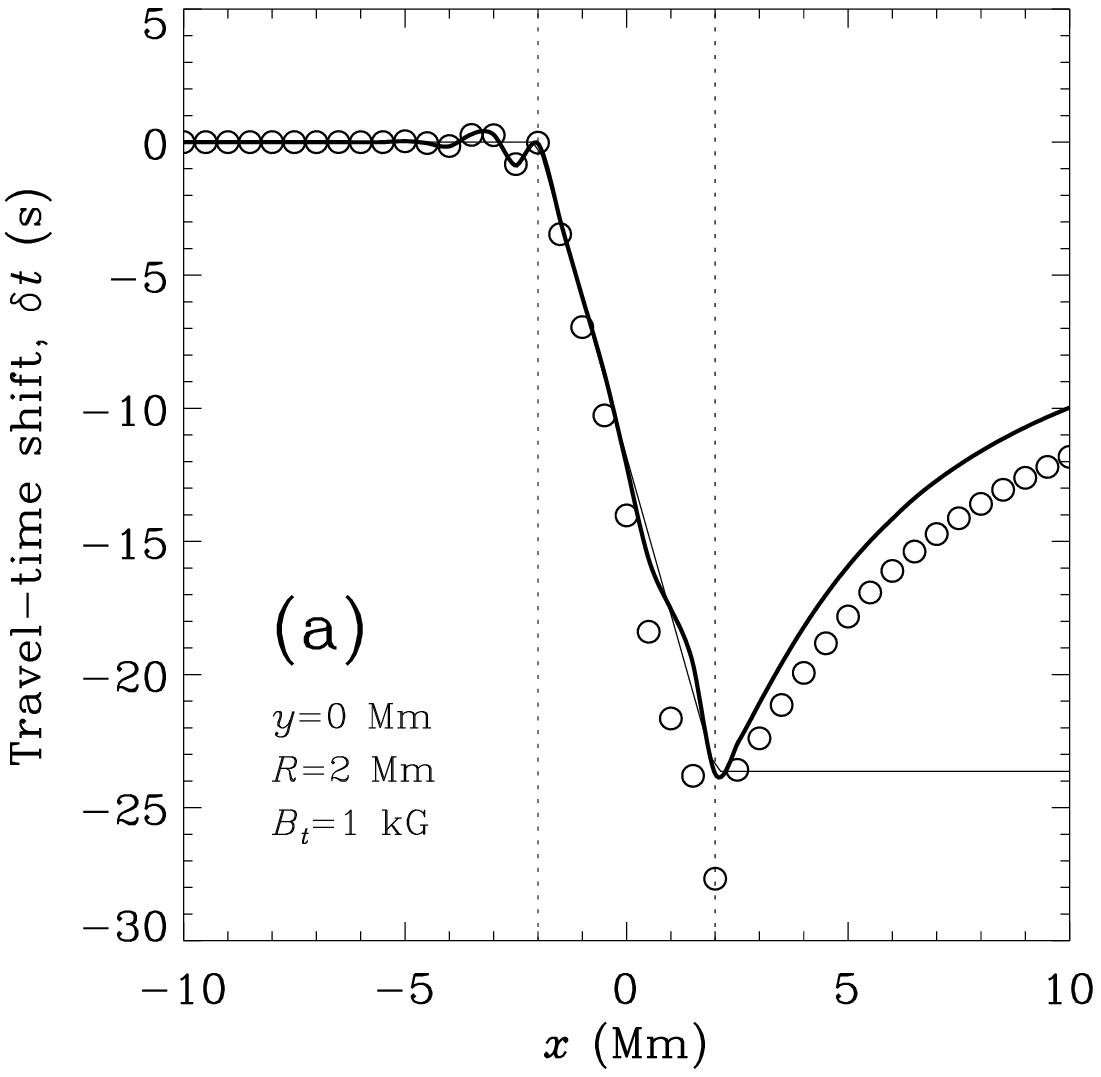}{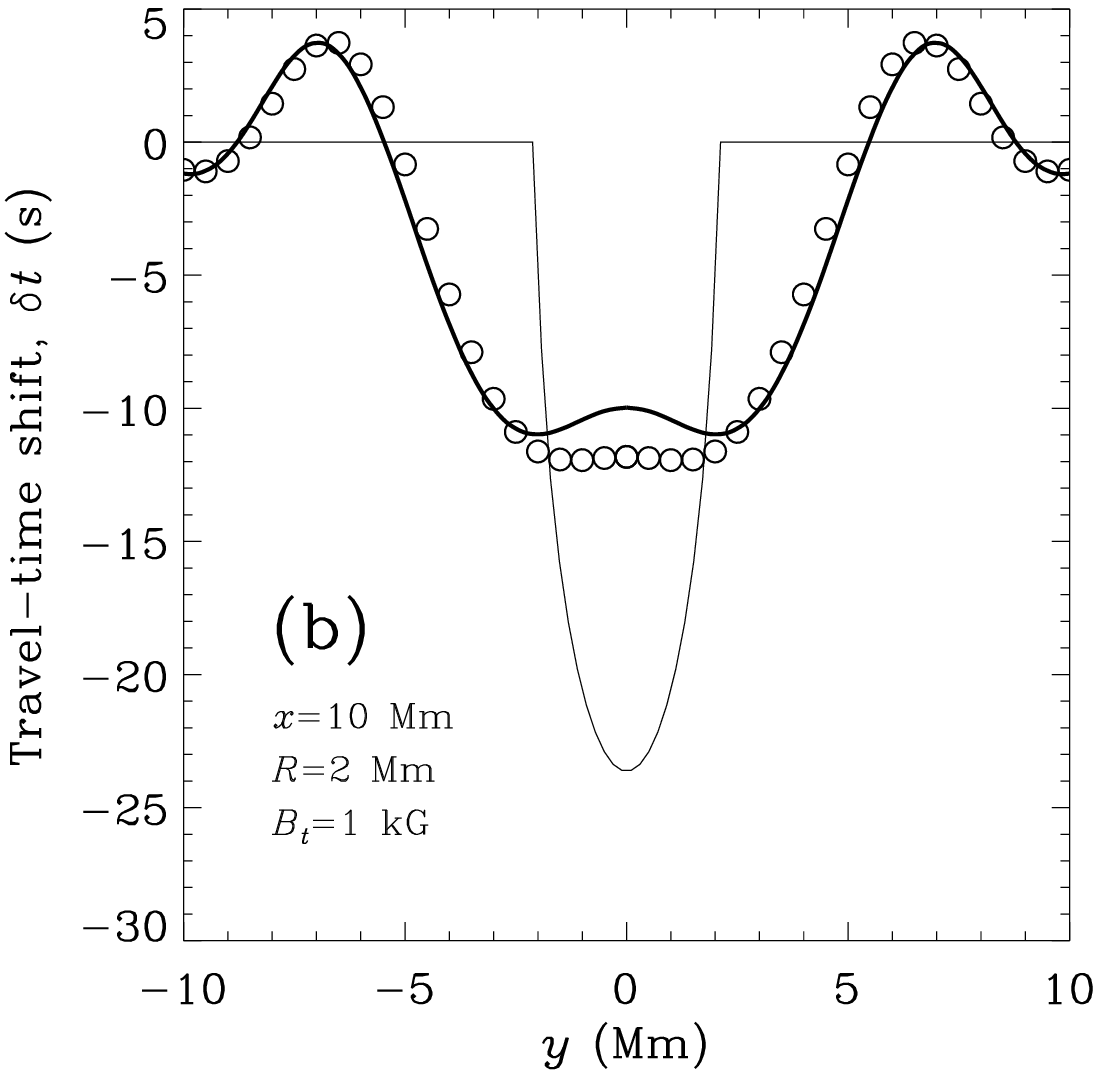}
\caption{Local travel-time shifts $\delta t(\br)$ caused by the magnetic cylinder ($\epsilon=0.13$). The travel times are  measured at  positions $\br$ in a plane perpendicular to both the cylinder axis. The incoming wavepacket, which moves in the $+\hat{\bf x}$ direction, is the same as in figure~\ref{fig.frames}a. The radius of the tube is $R=2$~Mm and the tube axis is $(x,y)=(0,0)$, as shown in figure~\ref{fig.frames}c.  In both panels the heavy solid line is the exact travel-time shifts, the circles are the Born travel-time shifts, and the light line gives the ray approximation. The left panel shows the travel-time shifts as a function of $x$ at fixed $y=0$.  The right panel shows the travel-time shifts as a function of $y$ at fixed $x=10$~Mm.  The Born approximation is reasonable for this value of $\epsilon$.  The ray-approximation does not capture finite-wavelength effects and fails to describe wavefront healing \citep{Nolet2000}.  }
\label{fig.travel_times}
\end{figure*}

Figure~\ref{fig.travel_times} shows travel-time shifts resulting from a flux tube of radius 2~Mm and field strength of 1~kG ($\epsilon=0.13$).  Figure~\ref{fig.travel_times}a shows the exact, Born-, and ray-approximation travel times as a function of $x$ at fixed $y$.  Inside the flux tube, both the Born- and ray-approximation travel times reproduce the exact travel times at a good level of accuracy.  As $x$ increases to the right of the tube, wavefront healing \citep[e.g.][]{Nolet2000} is seen in the exact and Born approximation travel times. Wavefront healing, however, is not seen in the ray approximation travel times. Figure~\ref{fig.travel_times}b shows the travel times as a function of $y$ at fixed $x=10$~Mm.  The Born approximation reproduces the exact travel times to within 20\%.  The ray approximation does not capture finite wavelength effects and does not capture the basic behavior of the travel times; it can be inaccurate by many orders of magnitude for $kR \ll 1$.  We note that, in Figure~\ref{fig.travel_times}, the contribution of the density jump (first term in Eq.~[\ref{source}]) to the travel-time shifts is negligible compared to the contribution from   the Lorentz force.

\section{Discussion}
\label{sec.discussion}
We have computed, in the first Born approximation, the scattering of acoustic waves from a magnetic cylinder embedded in a homogeneous background medium.  We showed that in the limit of weak magnetic field, the Born approximation to the scattered wavefield is correct to first order in the parameter $\epsilon=B^2/4\pi\rho c^2$.  For typical values of the solar magnetic flux, the Born approximation should be good  at depths larger than a few hundred km below the photosphere. The condition $\epsilon<<1$ is satisfied for a 1-kG magnetic fibril at a depth of 250~km ($\epsilon \approx 0.1$) and for a $10^5$~G magnetic flux tube at the base of the convection zone ($\epsilon \approx 10^{-7}$). 
Since the errors introduced by the Rytov and Born approximations are very similar \citep[e.g.][]{Woodward1989}, we suspect that a travel-time shift computed in the Rytov approximation would also tend to the exact solution as $\epsilon$ tends to zero.

Near the photosphere, $\epsilon$ is not small.  It has been suggested by many authors \citep[e.g.][]{Lindsey2004} that in this case the Born approximation will fail.  An exception is the claim by \citet{Rosenthal1995} that the Born approximation will remain valid for kG magnetic fibrils in the limit where the radius of the magnetic element is much smaller than the wavelength.
We wish to test this last statement in our simple problem.  

\begin{figure}
\plotone{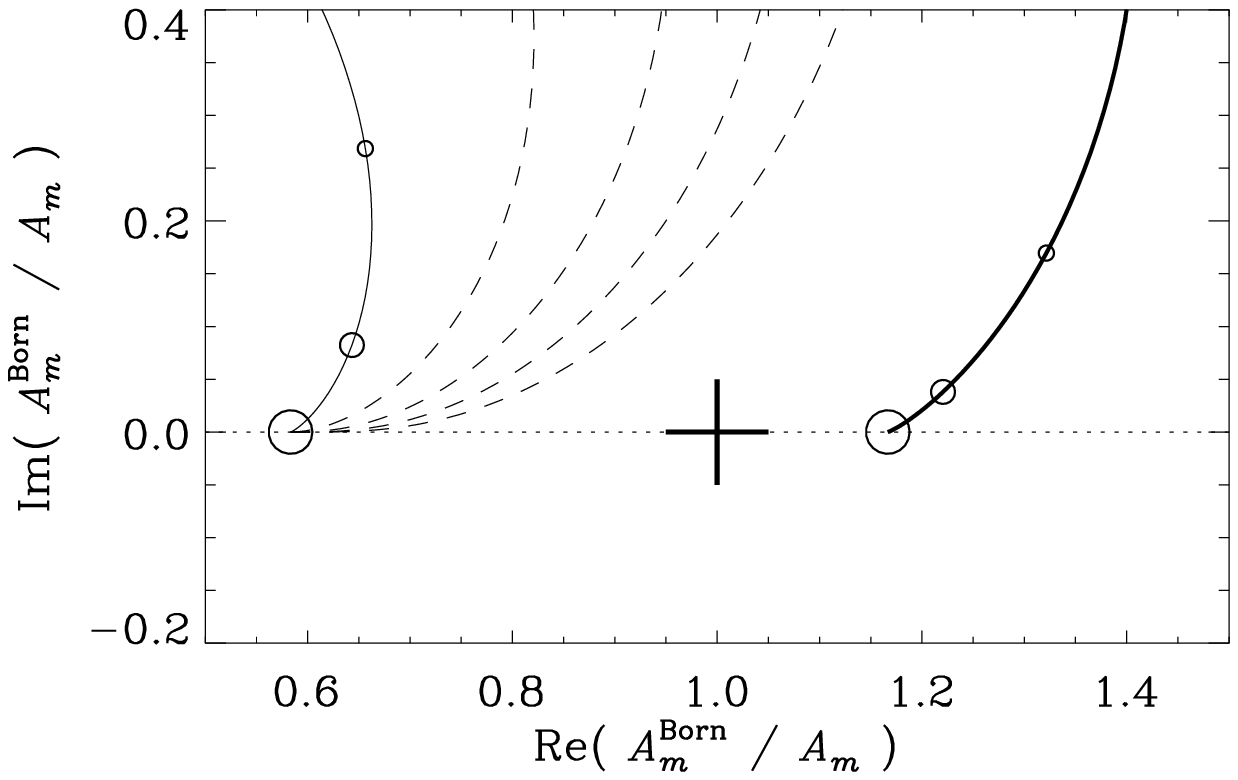}
\caption{Ratio $A^{\rm Born}_m/A_m$ in the complex plane at fixed $\epsilon=1$ and $k_z=0$. The ratio is plotted for varying values of the tube radius in the cases $m=0$ (thick line),  $m=1$ (thin line), and $2\leq m\leq 5$ (dashed lines).  The big circles show the limit $kR\rightarrow 0$ given by equation~(\ref{eq.ratio_Am}).  If the Born approximation were correct for small tube radii, the big circles would coincide with the cross. The small and medium-size circles are for $kR=1$ and $kR=1/2$ respectively.}
\label{fig.ratio_of_A}
\end{figure}

Assuming $k_z=0$ for the sake of simplicity and taking the limit $kR \rightarrow 0$, we find that for all $\epsilon$ we have
\begin{equation}
\lim_{kR \rightarrow 0} \frac{ A^{\rm Born}_m}{A_m}   = 
\left\{ 
\begin{array}{ll}
1 + (1-\gamma/2) \epsilon & {\rm if}\; m=0 , \\
1 - \gamma \epsilon / 4   & {\rm otherwise} .
\end{array}
\right.  
\label{eq.ratio_Am}
\end{equation}
This shows that the Born approximation is not valid in the limit of small tube radius.  Figure~\ref{fig.ratio_of_A} shows the ratio ${ A^{\rm Born}_m} / {A_m}$, for $0\leq m \leq 5$, as a function of $R$ when $\epsilon=1$ and $k=3.7$~Mm$^{-1}$.  We see that, in the limit of small $kR$, the fractional error in the Born approximation is of order $\epsilon$ (the absolute error is of order $\epsilon^2$).  The Born approximation applied to completely evacuated solar magnetic fibrils in the photosphere is likely to be invalid by roughly a factor of two. Note that the sign of the relative error in $A_m^{\rm Born}$ is different for $m=0$ and $m>0$.

The sensitivity of travel times to local perturbations in internal solar properties can be described through linear sensitivity functions, also called travel-time kernels. \citet{Gizon2002} gave a general recipe for computing  such travel-time kernels using the Born approximation, which has been applied to the case of sound-speed perturbations by \citet{Birch2004b}. The present work suggests that travel-time kernels for the subsurface magnetic field will be useful for probing depths greater than a few hundred km beneath the photosphere, at least in the case when the travel times are measured between surface points that are not in magnetic regions. One should be careful, however, not to draw definitive conclusions from the simple model we have studied, given the complexity of the real solar problem.

\begin{acknowledgements}
A.C.B. was supported by NASA contract NNH04CC05C and thanks the Max-Planck-Institut f\"{u}r Sonnensystemforschung for its hospitality.
\end{acknowledgements}


\appendix
\section{Exact solution coefficients}
\label{app.exact}
The coefficients $A_m$ and $B_m$ are 
\begin{equation}
A_m =   \frac{ - \left(1 + \frac{\gamma}{2}\frac{{a}^2}{c^2}\right)\frac{k_t}{k}J'_m({k_t R})J_m(kR) + \left(1 -  \frac{{a}^2 k_z^2}{\omega^2}\right)J_m({k_t R})J'_m(kR)}{\left(1 + \frac{\gamma}{2}\frac{{a}^2}{c^2}\right)\frac{k_t}{k}J'_m({k_t R})H_m(kR) - \left(1 - \frac{{a}^2 k_z^2}{\omega^2}\right)J_m({k_t R})H'_m(kR)}
\label{exact_out}
\end{equation}
and
\begin{equation}
B_m = -\frac{2i}{\pi kR}\left[\frac{k}{k_t}\left(1 + \frac{\gamma}{2}\frac{{a}^2}{c^2}\right)J'_m(k_t R)H_m(kR)-\frac{k^2}{k_t^2}\left(1 - \frac{{a}^2 k_z^2}{\omega^2}\right)J_m({k_t R})H'_m(kR)\right]^{-1}
\label{exact_in}  .
\end{equation}
 In equations~(\ref{exact_out}) and (\ref{exact_in}), the functions $J_m'$ and $H_m'$ denote the first derivative of $J_m$ and $H_m=H_m^{(1)}$ respectively.  
\section{Useful Integrals}
In order to compute scattering amplitudes in the Born approximation, 
we used \citep[][chap. 5]{Watson1944}
\begin{equation}
\int^x x'J^2_m(kx') \; d x' = \frac{x^2}{2}\left[J_m^2(kx) - J_{m-1}(kx)J_{m+1}(kx)\right] 
\label{eq.appendix1}
\end{equation}
and
\begin{equation}
\int^x x'H_m(kx')J_m(kx') \; d x' = \frac{x^2}{4}\left[2J_m(kx) H_m(kx) - J_{m-1}(kx) H_{m+1}(kx) - J_{m+1}(kx)H_{m-1}(kx)\right]  .
\label{eq.appendix2} 
\end{equation}

\section{Born approximation  coefficients}
\label{app.Born}
The coefficients $A_m^{\rm Born}$ and $C_m$ for the Born solution are
\begin{equation}
A^{\rm Born}_m = -\epsilon \frac{i\pi kR}{4}\left[\left(\gamma + 2\frac{c^2k_z^2}{\omega^2}\right)J'_m(kR)J_m(kR) + kR J_m^2(kR) - kR J_{m-1}(kR)J_{m+1}(kR) \right]   
\label{born_out}
\end{equation}
and 
\begin{eqnarray}
C_m &=& - \epsilon \frac{c^2k^2}{\omega^2}   -  \epsilon \frac{i\pi kR}{4}    
     \left(\gamma + 2 \frac{c^2k_z^2}{\omega^2}\right)J'_m(kR)H_m(kR)  \nonumber\\ && - \epsilon \frac{i \pi (kR)^2}{8}\left[2J_m(kR) H_m(kR) - J_{m-1}(kR) H_{m+1}(kR) - J_{m+1}(kR) H_{m-1}(kR)\right] .
\label{born_in}
\end{eqnarray}

\end{document}